# Harnessing Interlayer Magnetic Coupling for Efficient, Field-Free Current-Induced Magnetization Switching in a Magnetic Insulator

Leran Wang[1], Alejandro O. Leon[2*], Wenqing He[3], Zhongyu Liang[1], Xiaohan Li[3], Xiaoxiao Fang[1], Wenyun Yang[1], Licong Peng[4], Jinbo Yang[1,5*], Caihua Wan[3], Gerrit E. W. Bauer[6,7], Zhaochu Luo[1,5*]

[1]State Key Laboratory of Artificial Microstructure and Mesoscopic Physics, Institute of Condensed Matter Physics and Materials, School of Physics, Peking University, 100871 Beijing, China.
[2]Departamento de Física, Facultad de Ciencias Naturales, Matemática y del Medio Ambiente, Universidad Tecnológica Metropolitana, Las Palmeras 3360, Ñuñoa 780-0003, Santiago, Chile.
[3]Beijing National Laboratory for Condensed Matter Physics, Institute of Physics, University of Chinese Academy of Sciences, Chinese Academy of Sciences, 100190 Beijing, China.
[4]School of Materials Science and Engineering, Peking University, 100871 Beijing, China.
[5]Beijing Key Laboratory for Magnetoelectric Materials and Devices, 100871 Beijing, China
[6]Kavli Institute for Theoretical Sciences, University of the Chinese Academy of Sciences, 100864 Beijing, China.
[7]Advanced Institute for Materials Research (AIMR), Tohoku University, 980-8576 Sendai, Japan.

*Correspondence to: zhaochu.luo@pku.edu.cn (Z.Luo); aleonv@utem.cl (A.L.); jbyang@pku.edu.cn (J.Y.)

**Abstract:** Owing to the unique features of low Gilbert damping, long spin-diffusion lengths and zero Ohmic losses, magnetic insulators are promising candidate materials for next-generation spintronic applications. However, due to the localized magnetic moments and the complex metal-oxide interface between magnetic insulators and heavy metals, spin-functional Dzyaloshinskii–Moriya interactions or spin Hall and Edelstein effects are weak, which diminishes the performance of these typical building blocks for spintronic devices. Here, we exploit the exchange coupling between metallic and insulating magnets for efficient electrical manipulation of heavy metal/magnetic insulator heterostructures. By inserting a thin Co layer, we enhance the spin-orbit torque efficiency by more than 20 times, which significantly reduces the switching current density. Moreover, we demonstrate field-free current-induced magnetization switching caused by a symmetry-breaking non-collinear magnetic texture. Our work launches magnetic insulators as an alternative platform of low-power spintronic devices.

# 1. Introduction

Engineering the magnetic coupling plays a crucial role in determining the functionalities of spintronic devices. The interlayer Ruderman–Kittel–Kasuya–Yosida (RKKY) interaction[1] and exchange-bias effect[2] stabilize the magnetic reference layer[3,4] in the magnetic tunnel junctions of non-volatile magnetoresistive random-access memories (MRAM), while the intralayer dipolar and chiral couplings are essential for scalable spin logic[5-7] and neuromorphic computing devices[8-10]. Research on the physics and applications of these couplings focusses on electric conductors such as magnetic/non-magnetic and ferromagnetic/antiferromagnetic metal bilayers, but much of the corresponding phenomenology in magnetic insulators remains to be explored. Taking advantages of low Gilbert damping[11,12], long spin-diffusion lengths[13-15] and zero Ohmic losses, magnetic insulators, particularly the iron garnets, are promising candidate materials for low-power and high-speed spintronic applications. Pioneering studies on magnetic metal/magnetic insulator bilayers demonstrated injection and modulation of spin waves[16-20], but due to the complexity of polycrystalline metal-oxide interfaces[21-23] a full understanding of the underlying mechanisms remains elusive.

Here, we report experiments on a high-quality magnetic heterostructure comprising of an ultrathin Co film on an epitaxial terbium iron garnet ($Tb_3Fe_5O_{12}$, TbIG) layer, which shows a strong ferromagnetic exchange coupling of ~69.0 μJ/m$^2$. By growing a heavy metal layer on top of the Co/TbIG bilayer, we electrically detect and manipulate the TbIG magnetization via the direct and inverse spin Hall and/or Edelstein effects. The interlayer exchange coupling significantly enhances the interfacial spin-mixing conductance, leading to a large spin-Hall magnetoresistance (SMR) and a high spin-orbit torque (SOT) efficiency. Moreover, the non-collinear magnetization of the Co layer breaks the mirror symmetry of the device, thereby facilitating current-induced switching of the TbIG magnetization without external magnetic fields. Our work demonstrates an interplay between magnetic coupling and SOTs that can help to design reliable and efficient magnetic memory devices and pave a new pathway for magnetic insulator spintronics.

# 2. Structural and magnetic properties of Pt/TbIG

The insulating rare-earth iron garnet TbIG was epitaxially grown on (111)-oriented single-crystalline gadolinium gallium garnet ($Gd_3Ga_5O_{12}$, GGG) substrates by pulsed laser deposition (PLD) (see Methods). The structure of TbIG is constructed with three magnetic sublattices: an octahedral Fe, a tetrahedral Fe, and a dodecahedral Tb sublattice[24] (**Figure 1a**). TbIG films with various thicknesses ranging from 5 to 53 nm were deposited for different experiments. The 5 nm-thick TbIG was used for the electrical transport measurement, whereas the thick TbIG yielding a stronger structural signal was used for X-ray diffraction (XRD) analysis that gives insight into the microstructure and strain state. As shown in Figure 1b, we present the symmetric XRD scans of a 53 nm-thick TbIG around the (444) peak of GGG substrate. The rocking curve gives a full-width at half-maximum (FWHM) of ~0.02° (see Supplementary Information S1), indicating a high-quality single crystallinity. We can further extract the lattice constant of $d_{444}$ = 0.1815 nm for the epitaxial TbIG film which is slightly larger than that (~0.1795 nm) in a TbIG bulk[25], implying a strained state originating from the TbIG/GGG interface[26,27]. We then conducted the magnetization measurements using vibrating-sample magnetometer (VSM) and magneto-optical Kerr microscopy (MOKE) to study the magnetic properties of TbIG films (Figure 1c). The magnetization hysteresis loops with out-of-plane magnetic fields reveal a strong perpendicular magnetic anisotropy and a small saturation magnetization ($M_S$ ~23 emu/cc) as a result of the partial compensation of sublattice magnetizations.

We sputtered a 5 nm-thick Pt layer on TbIG (5 nm)/GGG with a low deposition power (20 W) to mitigate the bombardment damage of Pt atoms on the TbIG surface. The cross-sectional high-resolution transmission electron microscopy (HR-TEM) image confirms the clean and sharp interface in the heterostructure that is prerequisite to efficiently manipulate the magnetization by spin currents (Figure 1d). The Pt/TbIG heterostructure was then patterned into Hall bar structures with the channel width of 10 μm using UV photolithography combined with the ion milling process. We measured the electrical transport with out-of-plane magnetic fields and observed an anomalous Hall resistance of -1.3 mΩ at room temperature (Figure 1e). By varying the measurement temperature, the coercivity exhibits a peak at a critical temperature at which the polarity of the anomalous Hall resistance is reversed, indicating magnetic moment compensation in TbIG around 270 K ($T_{comp}$) (Figure 1f). Since the sign of the anomalous Hall resistance depends on the direction of the magnetic moment of the tetrahedral Fe sublattice[28], the sign is reversed when the temperature varies across the compensation state due to the fact that the alignment relationship between the magnetic field and the magnetic moment of the tetrahedral Fe sublattice reverses at $T_{\text{comp}}$ (Figure 1e).

**3. Spin Hall magnetoresistance and magnetic coupling in Pt/Co/TbIG**

To exploit the rich interfacial magnetic effect of heavy metal/metallic magnet such as high spin-mixing conductance and large Dzyaloshinskii–Moriya interaction (DMI), we inserted a thin Co layer ($t_{\text{Co}}$ = 0.3, 0.6 and 0.9 nm as calculated by multiplying the deposition rate and time) between the Pt and TbIG layers by DC magnetron sputtering (**Figure 2a**). With the insertion of Co layers, a linear Hall resistance with positive slope emerges on top of the square-like anomalous Hall loop, indicating an in-plane magnetic anisotropy of the Co layer (Figure 2b). The Co magnetization gets fully perpendicular at large out-of-plane magnetic fields (Figure 2c). Interestingly, the magnitude of the square-like anomalous Hall loop is significantly enhanced for thicker Co films, accompanying a change of sign of the Hall resistance from negative to positive (Figure 2d). Moreover, the anomalous Hall loop in Pt/Co (0.9 nm)/TbIG of 52.0 mΩ is more than an order of magnitude larger than that in Pt/TbIG (-1.3 mΩ) with the same TbIG thickness. An ultrathin Co insertion therefore offers a sensitive method to electrically detect the magnetization direction of insulators.

Two mechanisms may cause the large anomalous Hall loop. On one hand, the exchange coupling between Co and TbIG may force a non-collinear alignment of the Co and TbIG magnetizations such that the out-of-plane component of the Co magnetization contributes to the anomalous Hall effect. In this scenario, the longitudinal ($R_{xx}$) and transverse Hall ($R_{xy}$) resistance can be written as:

$$R_{xx} = R_0 + \Delta R_{\text{AMR}} \sin^2 \theta_{\text{Co}} \sin^2 \varphi_{\text{Co}}, \tag{1}$$

$$R_{xy} = -\Delta R_{\text{AMR}} \sin^2 \theta_{\text{Co}} \sin 2\varphi_{\text{Co}} + R_{\text{AHE}}^{\text{Co}} \cos\theta_{\text{Co}} + R_{\text{OHE}} H_z, \tag{2}$$

where $R_0$ and $\Delta R_{\text{AMR}}$ represent the magnetization-independent longitudinal resistance and the anisotropic magnetoresistance (AMR), respectively. $R_{\text{AHE}}^{\text{Co}}$ ($R_{\text{OHE}}$) is the anomalous (ordinary) Hall resistance. $\theta_{\text{Co}}$ ($\varphi_{\text{Co}}$) is the angle between Co magnetization and z-axis (y-axis) as defined in Figure 2a. We disregard the very small magnetic-proximity effect in Pt (see Supplementary Information S2). The second mechanism would follow the spin Hall magnetoresistance (SMR) scenario as widely used to extract spin transport parameters in heavy metal/magnetic insulator bilayers[29,30] such as the interfacial spin-mixing conductance $G$. In such measurements, an electric current is applied to the heavy metal layer and due to the spin Hall effect, it can generate a transverse spin current that will either be transmitted or reflected at the heavy metal/magnetic insulator interface depending on the relative orientations between the spin current polarization $\sigma$ and the magnetization $m$. The ratio of spin transmission/reflection at the interface will further modulate the electric current in the heavy metal layer due to the inverse spin Hall effect. This leads to the presence of a resistance change that has a distinct angle dependence. In the high field limit, the magnetizations of Co and TbIG are aligned along the field direction with angle $\theta$ ($\varphi$) with the $z$-axis ($y$-axis). The longitudinal ($R_{xx}$) and transverse Hall ($R_{xy}$) resistance then read[29,30]:

$$R_{xx} = R_0 - \Delta R_{\text{SMR}} \sin^2\theta \cos^2\varphi, \tag{3}$$

$$R_{xy} = -\Delta R_{\text{SMR}} \sin^2\theta \sin 2\varphi + R_{\text{AHE}}^{\text{SMR}} \cos\theta + R_{\text{OHE}}H_z, \tag{4}$$

where $\Delta R_{\text{SMR}}$ and $R_{\text{AHE}}^{\text{SMR}}$ are the SMR and the SMR-induced anomalous Hall resistances of the TbIG/Co interfaces adjacent to Pt, respectively, which are a function of the Co thickness.

Note that the transverse SMR and the SMR-induced anomalous Hall effect (AHE) have the same symmetry as the AMR-induced transverse (planar Hall) resistance and anomalous Hall resistance in Co, respectively. Hence, we cannot quantitatively distinguish the contribution of anomalous Hall resistance from the SMR-induced AHE and conventional AHE by the transverse resistance only (see Supplementary Information S3). By contrast, the contributions of the AMR and SMR to the longitudinal resistance changes distinctly depend on the magnetization angles and can be resolved by angular-dependent measurements. As shown in Figures 2e-2g, we measured $R_{xx}$ with a rotating magnetic field of 60 kOe in $zx$, $zy$ and $xy$ planes. According to the symmetry of SMR and AMR, a fit to the angular-dependent data in $zy$ plane yields $\Delta R_{\text{SMR}}$ = 61.1, 87.7, 225.9 and 555.7 mΩ for different Co thickness of $t_{\text{Co}}$=0, 0.3, 0.6 and 0.9 nm, respectively, while a fit to the angular-dependent data in $zx$ plane yields $\Delta R_{\text{AMR}}$ = -0.9, 1.4, 13.2 and 72.6 mΩ. The sizeable $\Delta R_{\text{SMR}}$ implies a substantial absorption of the spin currents by the heavy metal/magnets interface.

Interfacial spin transmission/reflection can be quantified by the complex spin-mixing conductance $G$ with a real ($G_r$) and an imaginary ($G_i$) part[29,30]. $G_r$ is related to the damping-like torques acting upon $\boldsymbol{m}$, which is proportional to $\boldsymbol{m}\times(\boldsymbol{m}\times\boldsymbol{\sigma})$ and manifests itself as a longitudinal SMR resistance $\Delta R_{\mathrm{SMR}}$. On the other hand, the imaginary part $G_{\mathrm{i}}$ is associated with a field-like torque, which is proportional to $\boldsymbol{m}\times\boldsymbol{\sigma}$ and contributes a Hall resistance $R_{\mathrm{AHE}}^{\mathrm{SMR}}$ in Pt. The experimental values of $\Delta R_{\mathrm{SMR}}$ and $R_{\mathrm{AHE}}^{\mathrm{SMR}}$, and assuming spin Hall angle $\theta_{\mathrm{Pt}} = 0.08$ and spin diffusion length $\lambda_{\mathrm{Pt}} = 1.4$ nm for Pt[30,32], we can estimate $G_{\mathrm{r}} = 4.3\times10^{14}$ $\Omega^{-1}\mathrm{m}^{-2}$ and $G_{\mathrm{i}} = -1.2\times10^{13}$ $\Omega^{-1}\mathrm{m}^{-2}$, which is consistent with reported values for Pt/magnetic garnet bilayers[30-32]. The insertion of the Co layer greatly enhances the spin-mixing conductance up to one order of magnitude to a $G_{\mathrm{r}}$ for $t_{\mathrm{Co}}$ of 0.6 and 0.9 nm of $3.9\times10^{15}$ $\Omega^{-1}\mathrm{m}^{-2}$ and $-5.8\times10^{15}$ $\Omega^{-1}\mathrm{m}^{-2}$, respectively, much larger than that in simple Pt/TbIG or Pt/Co heterostructures[33], implying that the exchange coupling plays a critical role in the spin-dependent transport.

To unravel the effect of the Co spacer, we recorded the hysteresis loop by sweeping the in-plane magnetic field. In Pt/TbIG, the Hall resistance exhibits the usual shape for a perpendicular magnetization under an in-plane magnetic field (**Figure 3a**). By fitting the in-plane hysteresis loop, we can obtain the effective magnetic anisotropy field $H_{\mathrm{K}} = 8$ kOe and magnetic anisotropy energy $E_{\mathrm{an}} = H_{\mathrm{K}}M_{\mathrm{S}}/2 = 9.2\times10^3$ J m$^{-3}$ (see Supplementary Information S6). In contrast, Pt/Co (0.9 nm)/TbIG shows a distinctly different shape with a sharp peak at small magnetic fields and saturation at a higher magnetic field (Figure 3b). This curve contains information about the exchange coupling parameter $J$ between Co and TbIG. The magnetic energy per unit area of Pt/Co/TbIG can be written as:

$$E = -J\hat{\boldsymbol{m}}_{\mathrm{Co}} \cdot \hat{\boldsymbol{m}}_{\mathrm{TbIG}} - K_{\mathrm{TbIG}}t_{\mathrm{TbIG}}\hat{m}_{\mathrm{TbIG},z}^2 - K_{\mathrm{Co}}t_{\mathrm{Co}}\hat{m}_{\mathrm{Co},z}^2 - \boldsymbol{H}\cdot(\boldsymbol{m}_{\mathrm{Co}} + \boldsymbol{m}_{\mathrm{TbIG}}), \quad (5)$$

where $K_{\mathrm{Co(TbIG)}}$ is the magnetic anisotropy coefficient and $\hat{\boldsymbol{m}}_{\mathrm{Co(TbIG)}}$ is the unit vector of the magnetization $\boldsymbol{m}_{\mathrm{Co(TbIG)}}$ of Co (TbIG). At zero magnetic field, the Co and TbIG magnetization unit vectors $\hat{\boldsymbol{m}}_{\mathrm{Co}}$ and $\hat{\boldsymbol{m}}_{\mathrm{TbIG}}$ are in-plane and out-of-plane, respectively. Minimizing Eq. (5) for $H=0$ and leading order in the interface exchange $J$ leads to:

$$\hat{m}_{\mathrm{Co},z} = \pm J/(2|K_{\mathrm{Co}}|t_{\mathrm{Co}}),$$
$$\hat{m}_{\mathrm{TbIG},x} = J/(2|K_{\mathrm{TbIG}}|t_{\mathrm{TbIG}}), \quad (6)$$

where the sign of $\pm$ holds for up- and downward TbIG magnetization (see Supplement S4). A relatively weak in-plane magnetic field overcomes the exchange coupling and pulls the Co magnetization into the plane, leading to a sharp decrease of the Hall resistance around zero magnetic field (Figure 3c). Due to its large perpendicular magnetic anisotropy, the TbIG magnetization persists to be perpendicular up to higher in-plane magnetic fields. The gradual pull into the plane is accompanied by a reduced and ultimately vanishing Hall resistance. The effective magnetic anisotropy field $H_\text{K}$ = 53.0 kOe of TbIG in Co (0.9 nm)/TbIG is even larger than that in pure TbIG (see Figure 3b). It has been reported that the interfaces of Pt/Co and Co/oxide can give large interfacial perpendicular magnetic anisotropy due to spin-orbit coupling[34]. Owing to the interlayer exchange coupling and proximity effect, the perpendicular magnetic anisotropy of TbIG may get enhancement from Pt/Co interfaces. We can then deduce a Hall resistance caused by the out-of-plane tilt of the Co magnetization ($R_\text{AHE}^\text{Co}$ = 71.0 mΩ), which is larger than the high-field TbIG/Co SMR ($R_\text{AHE}^\text{SMR}$ = -19.0 mΩ). We hence estimate $G_\text{i}$ in Pt/Co (0.9 nm)/TbIG to be -7.2×10$^{14}$ Ω$^{-1}$ m$^{-2}$, which is 60 times larger than that in Pt/TbIG. From the ratio of $R_\text{AHE}^\text{Co}$ at zero and that at the saturation out-of-plane magnetic fields (Figure 2c) the competition between Co/TbIG exchange coupling and magnetic anisotropies leads to a tilt angle of the Co magnetization of 83.2° at zero magnetic fields, *i.e.* pulled out of the plane by 6.8°. Substituting the observed magnetic anisotropies into the macrospin model, we arrive at an exchange coupling strength between Co and TbIG of $J$ = 69.0 μJ m$^{-2}$ (see Supplementary Information S4).

**4. Efficient current-induced magnetization switching**

The enhanced spin-mixing conductance implies efficient SOTs in current-biased Pt/Co/TbIG structures. We measured the SOT efficiency by recording magnetic hysteresis loops as a function of a charge current bias and in-plane and out-of-plane magnetic fields, a common technique for both metallic and insulating magnet/heavy metal bilayers[35-37]. As shown in **Figure 4a**, we can measure the damping-like SOTs by the current-induced shifts of the hysteresis loops since they give rise to out-of-plane effective fields $H_\text{eff}$ that act on the DMI-stabilized Néel-type domain walls[36]. We may define an SOT efficiency $\chi$ as:

$$\chi = H_\text{eff}/j. \quad (7)$$

Representative hysteresis loops of Pt/Co (0.6 nm)/TbIG with $H_x$ = 300 Oe and $j$ = ±4.3×10$^{10}$ Am$^{-2}$ are shown in Figure 4b. Changing the direction of the applied currents cause opposite shifts of the hysteresis loops as expected for a current-induced $H_\text{eff}$ caused by damping-like torques. The hysteresis loops shifts increase with an in-plane magnetic field and saturate at a critical value that is governed by the DMI[32,33,35] (Figure 4c). In Pt/TbIG, the SOT efficiency saturates to $0.9 \times 10^{-14}$ TA$^{-1}$m$^2$ at $H_x$ = 340 Oe. We can estimate the modulus of the effective DMI constant $|D| = \mu_0 M_s \Delta |H_x|$ from the saturation field $H_x$, where $\mu_0$ is the vacuum permeability and $\Delta = (A/K_u)^{1/2}$ the width of the domain wall. Here, $A$ = 2.3 pJ m$^{-1}$ is the exchange stiffness and $K_u$ is the anisotropy energy obtained from the experiment, leading to $|D_{Pt/TbIG}|$ =8.6 µJ m$^{-2}$ consistent with other Pt/magnetic garnet bilayers[35,36,38-41]. Interestingly, the insertion of a thin Co layer between TbIG and Pt significantly enhances the SOT efficiency as well as the effective DMI. For Pt/Co (0.6 nm)/TbIG $\chi$=24.0×10$^{-14}$ TA$^{-1}$m$^2$ at $H_x$ = 500 Oe, more than an order of magnitude larger than that of other heavy metal/magnetic insulator systems[35,42].

The model of magnetization reversal by the motion and annihilation of Neel domain walls implies that the spin-orbit torque is most efficient at the domain wall center (see Figure 4a) where the magnetizations of Co and TbIG are strictly parallel. We may therefore carry over the analysis of the SMR in the collinear limit of high magnetic fields to understand the observed enhancement of the damping-like torque. Here the Co capping layer enhances the effective spin mixing conductance $G_r$ from a small value for pure TbIG to the large one of Co. Since $G_r$ measures the absorption of the transverse spin current by the ferromagnet, this result directly explains the enhanced spin transfer torque. The interface exchange coupling subsequently communicates the torque to the TbIG which leads to the motion of the entire domain wall. The Co overlayer ensures non-chiral domain walls required for the magnetization reversal. An interesting subject for future research is an analysis of the Co-thickness dependence of the spin transfer, since partial trapping of the spins into quantum wells[43,44] and the associated multiple scattering at the TbIG/Co interface could additionally increase $H_{\text{eff}}$.

Since large damping-like torques in Pt/Co/TbIG should reduce the critical currents that switch the magnetization, we measure the changes in the transverse resistance under current pulses with a duration of 0.2 ms and a DC current of up to 1 mA corresponding to a current density of 2.5×10$^{10}$ Am$^{-2}$ in the same direction (see Supplementary Information S5). In the presence of an in-plane magnetic field of 400 Oe along the current direction, pulses with an amplitude up to $j$ = 6.1×10$^{10}$ Am$^{-2}$ change the sign of the transverse resistance which indicates a complete TbIG magnetization switching (Figure 4d). Reversing the direction of the in-plane magnetic field leads to an anti-clockwise hysteresis loop (Figure 4e), which agrees with the sign of the spin Hall angle of Pt[45]. Increasing the in-plane magnetic fields reduces the energy barrier of magnetization reversal and the critical switching current density of TbIG/Co (0.6 nm)/Pt from $6.9 \times 10^{10}$ to $5.0 \times 10^{10}$ Am$^{-2}$ (Figure 4f) half of that in Pt/TbIG ($1.2 \times 10^{11}$ Am$^{-2}$ at $H_x$ = 204 Oe). The suppression of the critical switching current density is less than expected from the enhanced SOT efficiency between Pt/Co/TbIG and Pt/TbIG deduced above, which suggests that Joule heating plays a role in the magnetization switching at high current densities[46,47].

## 5. Field-free magnetization switching

Current-induced switching of perpendicular magnetization without the need to apply magnetic fields is highly desirable in high-density magnetic memories. To this end, various schemes have been proposed in metallic systems, e.g., by breaking the mirror symmetry in asymmetric lateral designs[48,49] and non-collinear magnetic alignment[50,51]. These invoke complex devices incorporating multiple exotic materials that might be difficult to realize with magnetic insulators. Here we may take advantage of the exchange coupling between Co and TbIG that leads to a non-collinear magnetic texture that allows switching of a perpendicular magnetization by electric currents without the assistance of magnetic fields (**Figure 5a**).

As in Pt/Co (0.6 nm)/TbIG, we observed a full current-induced switching of magnetization in Pt/Co (0.9 nm)/TbIG with the same switching polarity cycle as function of the in-plane magnetic field sweeps (Figures 5b-5c). Taking the device out of the electromagnet set-up, where the residual magnetic field is less than 1.0 Oe, does not deteriorate the switching performance (Figures 5d-5e). The switching polarity is determined by the history of the applied in-plane magnetic fields. When initialized with a positive (negative) in-plane magnetic field of 1 kOe, the electric current can switch the magnetization with anti-clockwise (clockwise) polarity. To verify the reliability of current-induced field-free switching, we repeated measurements by applying alternating current pulses of $\pm 1.43 \times 10^{11}$ Am$^{-2}$ in the absence of magnetic fields (Figure 5f). The Hall resistance jumps between the resistance levels corresponding to up and down magnetizations after every pulse. The switching polarity reverses when the direction of the pre-set in-plane magnetic field changes from positive to negative, which is consistent with the current-driven hysteresis loops.

The SOT efficiency measurement in Pt/Co (0.9 nm)/TbIG led to several interesting observations (Figure 5g). First, in contrast to Pt/Co (0.6 nm)/TbIG, the magnitude of SOT efficiency in Pt/Co (0.9 nm)/TbIG saturates at a low magnetic field, implying the existence of an effective in-plane magnetic field due to the coupling with the in-plane magnetized Co layer. Second, the saturation SOT efficiency ($2.6\times10^{-14}$ $TA^{-1}m^2$) in Pt/Co (0.9 nm)/TbIG is lower than that ($24.0\times10^{-14}$ $TA^{-1}m^2$) in Pt/Co (0.6 nm)/TbIG, though the spin-mixing conductance in Pt/Co (0.9 nm)/TbIG is higher than that in Pt/Co (0.6 nm)/TbIG. However, a thicker Co layer also dissipates more spin currents, leading to a reduced SOT efficiency. Therefore, there is a trade-off of Co thickness to obtain the optimum SOT efficiency. The SOT efficiency can be further enhanced by optimizing the thickness of TbIG[52] and tuning its magnetization compensation state[53]. Moreover, there is a substantial non-zero SOT efficiency at zero magnetic fields and its sign depends on the history of sweeping in-plane magnetic fields, supporting the performance of field-free magnetization switching.

## 6. Conclusion

In summary, we exploit the magnetic coupling between metallic and insulating magnets, to efficiently manipulate the magnetism in an insulator with perpendicular magnetization. The itinerant conduction electrons in metallic magnets provide rich and strong spin-related interfacial effects in magnetic trilayers. By harnessing the magnetic coupling between metallic and insulating magnets, these interfacial effects can be imprinted into magnetic insulators, allowing for efficient electrical detection and manipulation of magnetism. In addition, coating the interface with a few metallic magnetic atoms significantly enhances the SOT efficiency as well as the DMI. Furthermore, we demonstrate the performance of field-free current-induced magnetization switching that results from symmetry-breaking non-collinear magnetic textures, paving the way for scalable magnetic memory devices. Therefore, our work offers a new avenue to engineer efficient spin memory and logic devices based on magnetic insulators.

**Methods**

*Growth of TbIG films*: TbIG thin films were deposited on 5 mm × 5 mm $Ga_3Gd_5O_{12}$ (111) single-side-polished substrates via pulsed laser deposition (PLD) at a laser fluence of ~1.4 $Jcm^{-2}$, and a target-to-substrate distance of ~6 cm. During deposition, the substrate temperature was heated to 800°C and the oxygen pressure was 30 mtorr. After deposition, annealing process was performed to promote the epitaxial growth of TbIG films with high quality and the cooling rate of the chamber was 20°C $min^{-1}$. Epitaxial growth of the films was confirmed via a high-resolution X-ray diffraction $2\theta$ scan of the (444) reflection. The thickness of thick TbIG films was determined by X-ray reflectometry, whereas the thickness of thin TbIG films (~5 nm) was calculated by the number of laser pulses.

*Device fabrication and measurement*: After the deposition of the TbIG thin film via PLD, metallic layers (such as Pt and Co/Pt) were deposited by DC magnetron sputtering at room temperature with a base pressure $<5 \times 10^{-8}$ torr. The deposition rate was 0.026 $nms^{-1}$ for Pt and 0.011 $nms^{-1}$ for Co. The thickness of Co and Pt layers was calculated according to the deposition rate. These multilayers were patterned into Hall bars using a combination of UV photolithography and Ar ion milling technique with lateral dimensions of 10 μm × 35 μm (width × length). The magnetic properties of TbIG films were measured by superconducting quantum interference device (SQUID) magnetometer. For the room-temperature electrical transport measurements, hysteresis loops were measured by using Keithley 2400/6221 source meter and 2182 nanovolt meter to apply currents and to measure Hall voltages, respectively. The temperature-dependent transport measurement, including AMR and SMR measurements were performed in a physical property measurement system (PPMS), which can provide the environment of different temperatures and high magnetic fields.

**Figures and figure captions**

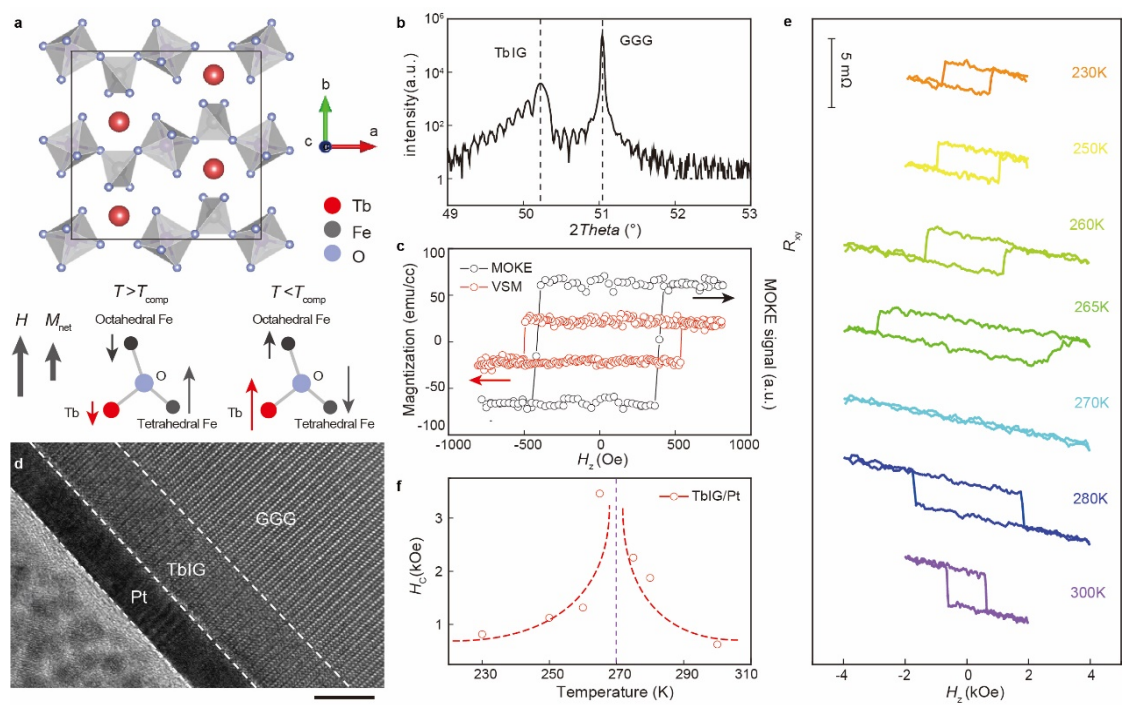

**Figure 1. Structural and magnetic properties of epitaxial TbIG films.** (**a**) Crystal structure of TbIG. The red, grey and blue balls represent the Tb, Fe and O atoms, respectively. Schematics of magnetic sublattices of TbIG illustrating the compensation state when the temperature is above or below the compensation temperature $T_{comp}$, are shown. (**b**) X-ray diffraction patterns ($2\theta$ scan) of a 53-nm-thick TbIG film grown on a (111)-oriented GGG substrate with peaks assigned to TbIG and GGG. (**c**) Magnetization as a function of out-of-plane magnetic field $H_z$ obtained by SQUID VSM (red curves) and MOKE (black curves) measurement. (**d**) Cross-sectional HR-TEM image of the TbIG/Pt heterostructure on a GGG substrate. Scale bar: 10 nm. (**e**) AHE hysteresis loops measured at different temperatures in Pt/TbIG. (**f**) Measured coercivity as a function of temperature measured in Pt/TbIG.

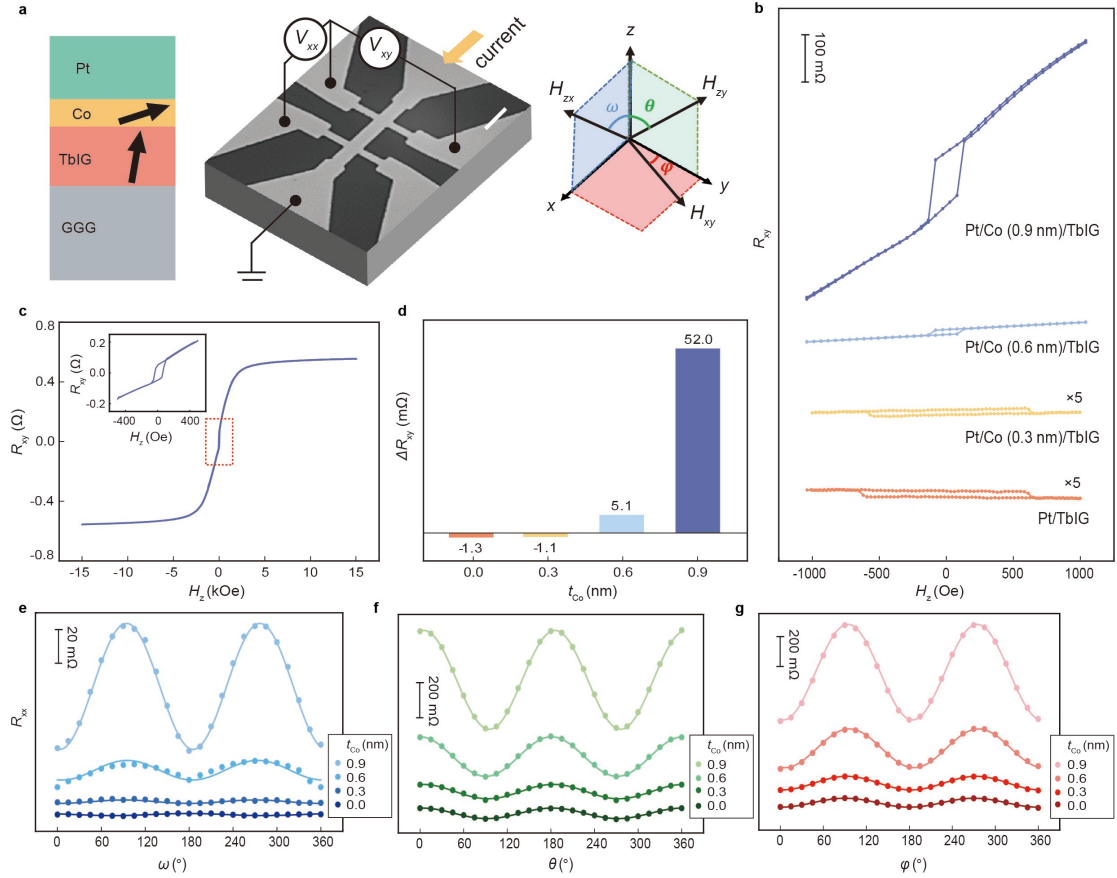

**Figure 2. Transport properties of Pt/Co/TbIG heterostructures.** (**a**) Schematics of Pt/Co/TbIG/GGG multilayer (not on scale) and optical image of a TbIG Hall bar device in a three-dimensional rendering of the measurement setup with white scale bar of 20 μm. The coordinate system and magnetic field angles are indicated. (**b**) AHE hysteresis loops measured in Pt/Co/TbIG for different thicknesses of the Co layer. (**c**) AHE hysteresis loops measured in Pt/Co (0.9 nm)/TbIG up to large out-of-plane magnetic fields. The AHE hysteresis loops at small magnetic fields are shown in the inset. (**d**) Magnitude of the anomalous Hall loops as a function of thickness of the Co layer. (**e-g**) Angular-dependent longitudinal resistance in Pt/Co/TbIG with an applied magnetic field of 60 kOe rotating in the *xy*, *xz* and *yz* planes.

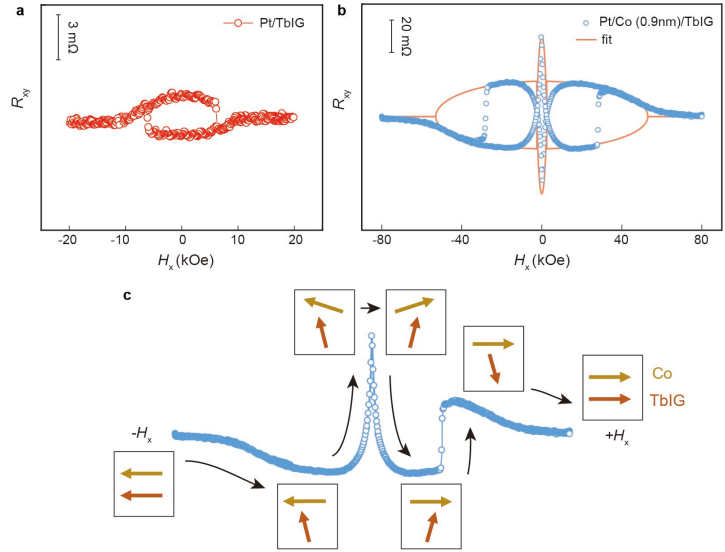

**Figure 3. AHE hysteresis loops with in-plane magnetic fields.** AHE hysteresis loops as a function of in-plane magnetic fields in (**a**) Pt/TbIG and (**b**) Pt/Co (0.9 nm)/TbIG. A fit to the data according to the experimental parameters and the model of magnetic coupling is indicated. (**c**) Schematics illustrating the magnetization when the in-plane magnetic field is swept from negative to positive in Pt/Co (0.9 nm)/TbIG.

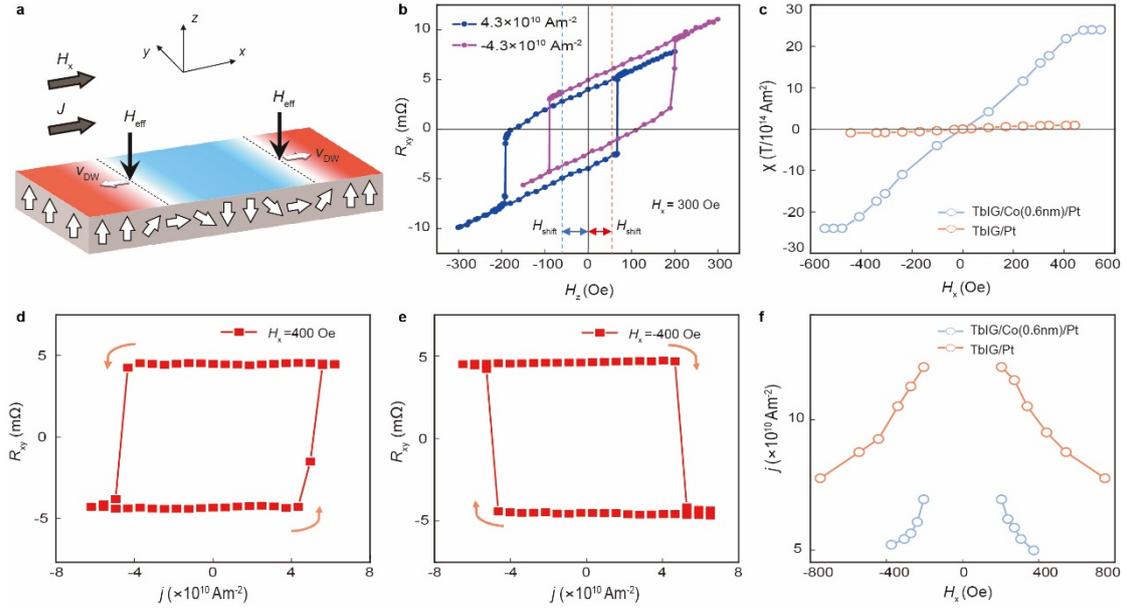

**Figure 4. Efficient current-induced SOTs in Pt/Co (0.6 nm)/TbIG.** (**a**) Current-induced domain-wall motion (domain expansion) with an in-plane magnetic field $H_x$ used to realign domain-wall moments. (**b**) AHE measurement with currents of $\pm 4.3\times 10^{10}$ Am$^{-2}$ under an in-plane magnetic field $H_x$ of 300 Oe. The horizontal shift of the hysteresis loops corresponds to the current-induced effective field $H_{\text{eff}}$. (**c**) SOT efficiency calculated from horizontal shifts for different current densities at different in-plane magnetic fields in Pt/TbIG and Pt/Co (0.6 nm)/TbIG. (**d**) and (**e**) Current-induced hysteresis loops with in-plane magnetic fields of $H_x = \pm 400$ Oe in Pt/Co (0.6 nm)/TbIG. The switching polarity is indicated by the arrows. (**f**) Critical switching current densities as a function of in-plane magnetic fields in Pt/TbIG and Pt/Co (0.6 nm)/TbIG.

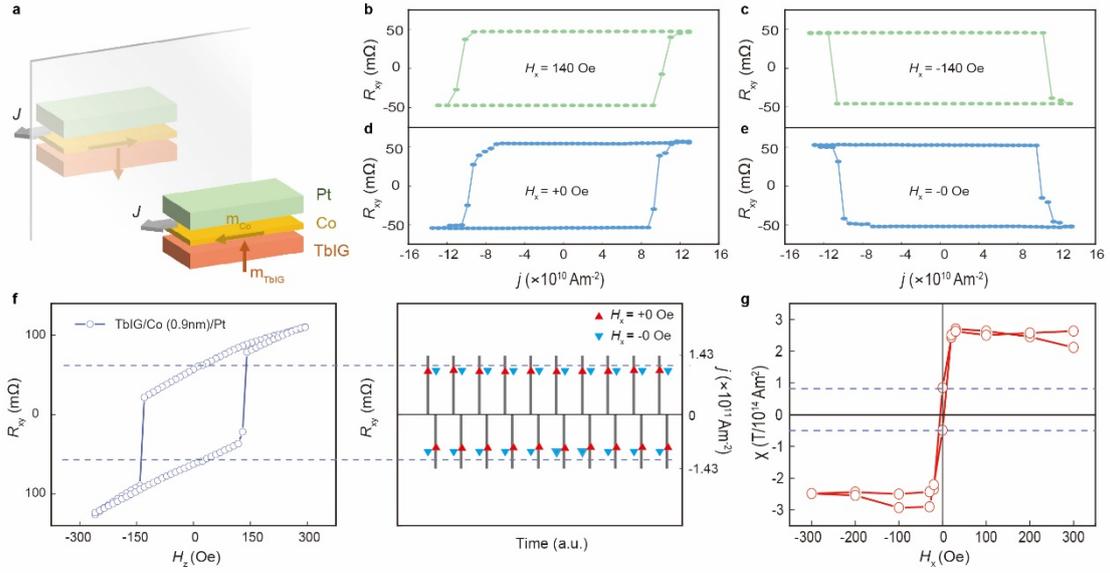

**Figure 5. Field-free magnetization switching in Pt/Co (0.9 nm)/TbIG.** (**a**) Mirror symmetry of the degenerate Pt/Co/TbIG magnetization configurations underlying field-free magnetization switching. (**b**) and (**c**) dc current-induced hysteresis loops with in-plane magnetic fields of $H_x = \pm 140$ Oe in Pt/Co (0.9 nm)/TbIG. (**d**) and (**e**): Current-induced hysteresis loops in the absence of magnetic fields in Pt/Co (0.9 nm)/TbIG after the pre-set with positive/negative in-plane magnetic fields of 1 kOe. (**f**) Current-induced magnetization switching with 0.2 ms-long current pulses of $\pm 1.4 \times 10^{11}$ Am$^{-2}$ in the absence of an external magnetic field after the pre-set with positive/negative in-plane magnetic fields of 1 kOe. The magnetic field-induced hysteresis loops are shown in left, giving the resistance reference for magnetizations pointing up and down. (**g**) SOT efficiency as a function of in-plane magnetic fields in Pt/Co (0.9 nm)/TbIG.

20